\magnification=\magstep1
\hfuzz=6pt
\baselineskip=16pt
\font\tiny=cmr8

$ $

\vskip 1in

\centerline{\bf When is a bit worth much more than $k_B T \ln 2$?}

\bigskip

\centerline{Can Gokler$^1$, Artemy Kolchinsky$^2$, Zi-Wen Liu$^3$,
Iman Marvian$^4$, Peter Shor$^5$,}
\centerline{ Oles Shtanko$^3$, Kevin Thompson$^1$,
David Wolpert$^{2,6,7}$,
Seth Lloyd$^{2,3,4,8,*}$}

\smallskip
{\tiny
\centerline{1. Harvard Engineering and Applied Sciences, 2. Santa Fe Institute, 3. MIT Physics,}
\centerline{ 4. MIT Research Laboratory of Electronics,
 5. MIT Mathematics, 6. MIT Aero Astro}
\centerline{7. Arizona State University, 8. MIT Mechanical Engineering}
\centerline{* to whom correspondence should be addressed: slloyd@mit.edu}}

\bigskip\noindent{\it Abstract:} Physical processes that
obtain, process, and erase information involve tradeoffs
between information and energy.    The fundamental energetic
value of a bit of information exchanged with a reservoir
at temperature $T$ is $k_B T \ln 2$.   This paper investigates
the situation in which 
information is missing about just what physical process is about
to take place.  The fundamental energetic value of such information
can be far greater than $k_B T \ln 2$ per bit. 

\vskip 1cm

Ever since Maxwell's introduction of his famous `demon' who
could use information to extract free energy from a system at
thermal equilibrium, it has been clear that physical processes
involve tradeoffs between energy, information, and
entropy [1-2].  Szilard's 1928 investigation of Maxwell's demon
in terms of a single-particle heat engine showed that the
fundamental energetic `value' of a bit of information exchanged with
a thermal reservoir at temperature $T$ is $k_B T \ln 2$, where
$k_B$ is Boltzmann's constant [3].    Similarly, Landauer's principle
[4] states that the erasure of a bit of information by interaction
with such a reservoir requires energy $k_B T\ln 2$.
In many macroscopic situations, however,
a bit of information can be worth far more that $k_B T\ln 2$.
For example, consider a situation in which an apple is hidden
in one of two boxes, and one is allowed to choose only one box,
and receive its contents.  If one knows which box the apple
is in (one bit of information), one can obtain the full energetic
value of the apple, say, 100 kilocalories.   If
one doesn't know which box the apple is in, then one can obtain
only 50 kilocalories on average.   The bit of information about
which box the apple is in is worth 50 kilocalories $\approx
10^{25} k_B T \ln 2$ at room temperature.   This anecdotal
example (there are many others) shows that information can
have far greater energetic value than $k_B T \ln 2$ per bit.
But are such macroscopic energetic values for bits fundamental?
Might there be a way of getting around the apparent wastefulness
of ignorance?  This paper provides a formal physical and
mathematical analysis to show that the answer that missing a bit
of information can fundamentally require macroscopic dissipation:
there is no way around it.  
We apply the Kolchinsky-Wolpert theorem [5] to
show that, generically, gaining a bit of
information about a macroscopic system
can allow one to gain large amounts of free energy.
Conversely, lacking that bit forces any attempt to harvest free
energy to undergo large amounts of dissipation.

The last few decades have seen a revolution in non-equilibrium
statistical mechanics [6-18], with the realization that many
thermodynamic processes are governed by exact and unexpected relations
such as the Jarzynski equality [6] and the Crooks fluctuation theorem [7].
The Kolchinsky-Wolpert theorem [5] is such an exact relation that
governs the amount of work dissipated
in an isothermal process.   It provides a simple formula that allows the
comparison between the minimum amount of work dissipated, and
the actual amount dissipated.  The K-W theorem states that a 
stochastic process in which a system exchanges energy and entropy
with a bath at temperature $T$, 
the excess dissipated work obeys
$$ W_D(r_0) - W_D(q_0) = k_B T\big( D(r_0\|q_0) - D(r_1\|q_1) \big).\eqno(1)$$
Here, $W_D(r_0)$ is the work dissipated into the environment
at temperature $T$ when the initial probabilities for the microstates
$x_0$ with energy $E_0(x_0)$ of the system are given by
$r_0(x_0)$; $W_D(q_0)$ is the minimum
work dissipated for the optimal initial probability distribution
$q_0(x_0)$ (under quite general conditions $q_0(x_0)$ has full support [5]);
$D(r\|q) = -\sum_x r(x) \ln \big(q(x)/r(x) \big)$ is the
Kullback-Leibler divergence/relative
entropy; $r_1(x_1)
= \sum_{x_0} r_0(x_0) p(x_1|x_0)$ are the output probabilities
when the input probability distribution was $r_0$;
$q_1(x_1)$ are the output probabilities given that the
input probability distribution was $q_0$.   The K-W theorem
is straightforward to derive and applies to arbitrarily
complicated stochastic processes.

In a companion paper [19] we show that the maximum increase in
the free energy of the system 
obeys a similar equation:
$$\Delta F(r_0) - \Delta F(\hat q_0) = 
- k_B T\big( D(r_0\|\hat q_0) - D(r_1\|\hat q_1) \big),
\eqno(2)$$
where $\Delta F(p_0)$ is the increase in free energy of the
system when the initial distribution over states is $p_0(x_0)$.
Here, $\hat q_0(x_0)$ is the initial probability distribution that
maximizes the increase in free energy over the process (which is
a different task from minimizing dissipated work, i.e.,
$\hat q_0 \neq q_0$ in general).
All of our results hold equally
for the maximum free energy increase as well as minimum dissipated
work.  In [19] we also show that these results hold for
quantum mechanical systems under the operation of completely
positive maps, with the quantum K-L divergence exchanged
for the classical K-L divergence.
  
We apply the K-W theorem 
to the fundamental problem raised in the introduction: When does
ignorance of the underlying stochastic process necessarily
lead to macroscopically large amounts of dissipation, or 
require one to forgo obtaining a large increase in free energy?

Consider the following situation.   We prepare
our system so that the initial probabilities for its microscopic
states $x_0$ are $r_0(x_0)$.  We then insert
our system into a `black box' in contact with a bath
at temperature $T$, where the system either undergoes stochastic
process $A$ or stochastic process $B$, each of which occurs with
probability $1/2$.    We don't know which process will occur in 
the box.   (Below, we generalize to more than two processes, occurring
with different probabilities.)
We minimize the excess dissipation over all initial
probability distributions $r_0$.    We then compare this minimum
dissipation in the absence of knowing which process takes place
with the minimum average dissipation $(1/2)( W_D(q_0^A) + W_D(q_0^B))$
that can be obtained if we do know which process takes place.
That is, we calculate the energetic value of the bit of information
that tells us whether the underlying process is $A$ or $B$.
From the K-W theorem, we see that
our goal is to find the initial distribution $r_0$ that minimizes
$$ \Delta =  (1/2) \big( D(r_0\|q_0^A) - D(r_1^A\| q_1^A)
+ D(r_0\|q_0^B) - D(r_1^B\| q_1^B) \big), \eqno(3)$$
where $r_1^{A,B}(x_1)$ are the final probabilities for $x_1$ when
the process in the box is $A,B$.

Let the two stochastic
processes that can be inside the box be defined by conditional
probabilities $p_A(x_1|x_0)$ and $p_B(x_1|x_0)$ 
for output states $x_1$ given input states $x_0$. 
The method of Lagrange multipliers yields an equation for the 
initial distribution $r_0(x_0)$ that minimizes the quantity
$\Delta$ in equation (3):
$$\eqalign{ & - \ln r_0(x_0) + (1/2) \big( 
\ln q_0^A(x_0) + \ln q_0^B(x_0) \big) \cr 
& + (1/2) \sum_{x_1} \big( \ln r_1^A(x_1) - \ln q_1^A(x_1) \big) 
p_A(x_1|x_0) \cr
& + (1/2) \sum_{x_1} \big( \ln r_1^B(x_1) - \ln q_1^B(x_1) \big) 
p_B(x_1|x_0) 
 = 0.\cr } \eqno(4)$$
The distribution $r_0$ that solves these equations is not obvious.
However, as we'll now show, the amount of excess dissipation can
easily be macroscopic.

As a simple example, suppose that the stochastic process $A$ 
always yields the same
final distribution $r_1^A(x_1)$, independent
of the initial distribution $r_0(x_0)$. 
Similarly, assume that the stochastic process $B$
always yields the same
final distribution $r_1^B(x_1)$, independent
of the initial distribution $r_0(x_0)$.
For example,  the process $A$ could always
end up with the system in a thermal state at temperature $T_A$,
while $B$ always ends up with a thermal state at temperature $T_B$.
In this setting, because the final state of the
process is ultimately the same no matter how one prepares the
initial state, the final relative entropy drops out of
equation (3), simplifying the calculation of the minimum 
dissipation.  Minimizing
the dissipation means finding the initial distribution 
$r_0$ that minimizes
$$ \Delta  = (1/2)\big( D(r_0\| q_0^A) + D(r_0 \| q_0^B ) \big) = 
D(r_0\| \sqrt{q_0^A q_0^B}).\eqno(5)$$
Note that $\sqrt{q_0^A(x) q_0^B(x)}$ is not in general
a probability distribution -- this is the key point -- but
it can still be inserted into the formula for the K-L divergence.

Because they are the same for all initial preparations of the state
of the system,
the final distributions $r_1^{A,B}$ now drop out of equation (3), and
a simple Lagrangian minimization shows that
the minimum dissipation occurs when
$$r_0 = e^{\hat \Delta} \sqrt{q_0^A q_0^B}.\eqno(6)$$
The minimum dissipation is $ k_B T \hat \Delta$, where
$$\hat \Delta 
= - \ln \big( ~ \sum_x \sqrt{q_0^A(x) q_0^B(x)} ~\big), \eqno(7)$$ 
to ensure that $r_0$ is a properly normalized
distribution.   In the quantum case, when the optimal
inital density matrices are $\chi_0^A$, $\chi_0^B$ for the
two quantum processes $A,B$, the minimum dissipation
occurs for initial density matrix $\rho_0$, where
$\ln \rho_0 = \hat \Delta + (1/2)( \ln \chi_0^A + \ln \chi_0^B)$.

The derivation of equations (6-7) immediately gives the generalization
to $K$ processes occurring with probability $p_k$.  Let the optimal
initial distribution for the $k$'th process be $q_0^k$.   Then
we have
$$r_0 = e^{\hat \Delta}
 \big(~ (q_0^1)^{p_1} \ldots (q_0^K)^{p_k} ~ \big),\eqno(8)$$
where
$$ \hat \Delta 
= - \ln \big(~ \sum_x (q_0^1(x))^{p_1} \ldots (q_0^k(x))^{p_k} ~ \big)
\big). \eqno(9)$$
Equations (6), (8) give the initial probability distribution
over microstates of the system that minimizes excess dissipation 
for processes that have a fixed final distribution.   

We now show that the excess dissipation
$k_B T \Delta$ can be -- and typically will be -- a
macroscopic quantity.    Consider thermodynamically
reversible processes $A$ and $B$, for which the minimum
excess dissipation is zero.   Suppose that as above,
$A$ takes any initial distribution to the same final distribution,
and $B$ does too (the final distribution for $B$ can be different
than the one for $A$).   In addition, suppose that $A$ is thermodynamically
reversible when the initial state of the system is a thermal
state
$$q_0^{A}(x_0) = { 1\over Z(\beta_{A}) } e^{-\beta_{A} E_0(x_0)}, \eqno(10)$$
where $\beta_A = 1/k_B T_A$.   Similarly, suppose that
$B$ is thermodynamically reversible for an initial thermal
state with the same energy function $E_0$, 
but a different inverse temperature $\beta_B = 1/k_B T_B$. 
For example, $A$ could be the optimal process for extracting work
from the initial thermal state $q_0^A$ by rapidly changing the system's energy
function/Hamiltonian,
putting the system in contact with a bath at temperature $T$, and
isothermally varying the Hamiltonian to its desired final form [16-18];
similarly, $B$ could be the optimal process for extracting work
from the initial thermal state $q_0^B$. 
Substituting these thermal distributions into equation (6)
shows that the optimal input distribution
$r_0$ is that of a thermal state with inverse temperature
$(1/2)(\beta_A + \beta_B)$, and equation (7)
yields
$$ \hat \Delta = (1/2) \big( \ln Z(\beta_A) + \ln Z(\beta_B) \big)
 - \ln Z\big( (1/2)(\beta_A + \beta_B) \big).\eqno(11)$$
Because the excess dissipation grows as the 
the number of subsystems grows, for macroscopic systems
$ \hat \Delta$ can become arbitrarily large.
When the system is macroscopic,
with $N$ subsystems, and $\beta_A$ and $\beta_B$ differ significantly,
the dissipation typically
grows proportionally to $N$.    For example,
the partition function $Z_N(\beta)$ for $N$ non-interacting, identical
systems, each with partition function $Z(\beta)$
grows as $Z_N(\beta) = (Z(\beta))^N$.   In 
this case, the excess dissipation $k_B T \hat \Delta$
grows as $ O(1) N k_B T$.   For interacting systems,
when $\beta_A$ and $\beta_B$ are
close in value, we can expand equation (10) in a Taylor series
around $(1/2)(\beta_A + \beta_B)$ to show that
the minimum excess dissipation is 
$$\hat \Delta = \langle (\Delta E)^2\rangle
(\beta_A - \beta_B)^2/2,\eqno(12)$$ 
where $\langle (\Delta E)^2\rangle$ is the
variance in energy, which also scales as $N$.

Equation (12) is a special instance of the general case
where the two processes $A$ and
$B$ are very similar, so that the optimal initial distributions
$q_0^A \approx q_0^B \approx q_0$.   Expanding equation
(3) to second order and minimizing yields excess
dissipation
$$\hat\Delta = (1/4) \big( \sum_{x_0} \delta q_0(x_0)^2/q_0(x_0)
- \sum_{x_1} \delta q_1(x_1)^2/q_1(x_1) \big), \eqno(13)$$
where $\delta q_0(x_0) = q_0^A(x_0) - q_0^B(x_0)$
and $\delta q_1(x_1) = q_1^A(x_1) - q_1^B(x_1)$.
That is, in the infinitesimal regime, the excess
dissipation is proportional to the 
Fisher information distance between the optimal
input distributions $q_0^A(x_0)$, $q_0^B(x_0)$,
minus the Fisher information distance between
the output distributions.  $\hat \Delta$ is non-negative
because of the data processing inequality for Fisher information [20]. 
In the quantum case the excess dissipation is proportional
to the Burres metric distance [21] between
input states minus the Burres distance between output states. 

\bigskip\noindent{\it Discussion:}  The Kolchinsky-Wolpert theorem
quantifies the amount of excess dissipation that occurs during
stochastic processes if one prepares
a physical system in the `wrong' initial state, i.e., a state that fails
to minimize dissipation.   Similarly, [19] quantifies the lost
free energy gain  when one prepares a physical system in
the wrong state.   This paper applied these 
results to the case where one is ignorant of the underlying
dynamics of the system.    Intuitively, if one doesn't know
what is going to happen, one's best efforts can be far more wasteful
than if one does know what is going to happen.   This paper provided
a rigorous treatment of this intuition in the case of stochastic
processes: even if one lacks only a single bit of information about
which stochastic process is going to take place, the best one
can do in extracting free energy and minimizing dissipation
can be far worse than if one possesses that bit.   

\vfill
{\it Acknowledgements:} The authors thank Juan Parrondo and Jordan 
Horowitz for helpful conversations.   This work was supported by NSF under an
INSPIRE program.  S.L. was supported by ARO and AFOSR.
AK and DHW would like to thank the Santa Fe Institute for helping
to support this research.  This paper was made possible through
the support of Grant No. TWCF0079/AB47 from the Templeton
World Charity Foundation, Grant No. FQXi-RHl3-1349 from the
FQXi foundation, and Grant No. CHE-1648973 from the U.S. National
Science Foundation.  The opinions expressed in this paper
are those of the authors and do not necessarily
reflect the view of Templeton World Charity Foundation.

\vfil\eject

\noindent{\it References}
\bigskip

\noindent [1] H.S. Leff, A.F. Rex, eds., 
{\it Maxwell's Demon: Entropy, Information, Computing,} 
Adam Hilger, Bristol (1990).

\smallskip\noindent [2] H.S. Leff, A.F. Rex, eds.  
{\it Maxwell's Demon 2: Entropy, Classical and Quantum Information, Computing,}
CRC Press, Boca Raton (2002).

\smallskip\noindent [3] Szilard, Leo, {\it Zeitschrift für Physik} 
{\bf 53}, 840–856 (1929).  (Reprinted in [1])

\smallskip\noindent [4] Landauer, Rolf, {\it IBM Jour.of Research and 
Development} {\bf 5}(3), 183–191 (1961). (Reprinted in [1]).

\smallskip\noindent [5] A. Kolchinsky, D.H. Wolpert,
`Dependence of dissipation on the initial distribution over states,'
arXiv: 1607.00956v2 (2016). 

\smallskip\noindent [6] C. Jarzynski, {\it Phys. Rev. Lett.} {\bf 78},
2690 (1997).

\smallskip\noindent [7]  G.E. Crooks, {\it Phys. Rev. E} {\bf 60}, 2721 (1999).

\smallskip\noindent [8]  G.E. Crooks, {\it J. Stat. Phys. 90,} 1481
(1998).

\smallskip\noindent [9]  H. Touchette and S. Lloyd, {\it Physica A}
{\bf 331}, 140 (2004).

\smallskip\noindent [10]  T. Sagawa and M. Ueda, {\it Phys. Rev. Lett.}
{\bf 102}, 250602 (2009).

\smallskip\noindent [11]  R. Dillenschneider and E. Lutz,  {\it Phys. Rev.
Lett.} {\bf 104}, 198903 (2010).

\smallskip\noindent [12]  M. Esposito and C. Van den Broeck,
{\it Phys. Rev. E}  {\bf 82}, 011143 (2010).

\smallskip\noindent [13] K. Wiesner, M. Gu, E. Rieper, and V. Vedral,
{\it Proc.  Roy. Soc. A} {\bf 468}, 4058 (2012).

\smallskip\noindent [14] T. Sagawa and M. Ueda, {\it Phys. Rev. Lett.} {\bf 109},
 180602 (2012).

\smallskip\noindent [15] S. Still, D. A. Sivak, A. J. Bell, and G. E. Crooks,
{\it Phys. Rev. Lett.} {\bf 109}, 120604 (2012).

\smallskip\noindent [16] J.M.R. Parrondo, J.M Horowitz, T. Sagawa,
{\it Nat. Phys.} {\bf 11}, 131-139 (2015).

\smallskip\noindent [17]  H-H. Hasegawa, J. Ishikawa, K. Takara, D.J. Driebe, 
{\it Phys. Lett. A} {\bf 374}, 1001–1004 (2010).

\smallskip\noindent [18] K. Takara, H-H. Hasegawa, D.J. Driebe, 
{\it Phys. Lett. A} {\bf 375}, 88–92 (2010).

\smallskip\noindent [19] A. Kolchinksky, I. Marvian, C. Gokler, Z.-W. Liu,
P. Shor, O. Shtanko, K. Thompson, D. Wolpert, S. Lloyd,
`Maximizing free energy gain,' arXiv: 1705.00041.

\smallskip\noindent [20] R. Zamir, {\it IEEE Info. Th.} {\bf 44}, 1246-1250
(1998).

\smallskip\noindent [21] M.M. Wilde, {\it Quantum Information Theory,}
Cambridge University Press, Cambridge (2013).

\vfill\eject\end